\documentclass[11pt,twoside]{article}
\usepackage{newpasp}

\def\edcomment#1{\iffalse\marginpar{\raggedright\sl#1\/}\else\relax\fi}
\reversemarginpar

\begin{document}
\title{Dark Matter in Galaxies: Conference Summary}
 \author{James Binney}
\affil{Theoretical Physics, Oxford University, 1 Keble Road, Oxford OX1 3NP,
England}

\begin{abstract}
The competition between CDM and MOND to account for the `missing mass'
phenomena is asymmetric. MOND has clearly demonstrated that a characteristic
acceleration $a_0$ underlies the data and understanding what gives rise to
$a_0$ is an important task. The reason for MOND's success may lie in
either the details of galaxy formation, or an advance in fundamental
physics that reduces to MOND in a suitable limit.  CDM has enjoyed great
success on large scales.  The theory cannot be definitively tested on small
scales until galaxy formation has been understood because baryons either
are, or possibly have been, dominant in all small-scale objects. MOND's
predictive power is seriously undermined by its isolation from the rest of
physics. In view of this isolation, the way forward is probably to treat CDM
as an established theory to be used alongside relativity and
electromagnetism in efforts to understand the formation and evolution of
galaxies.
\end{abstract}

\section{Introduction}

In the widely accepted Popperian interpretation of the scientific process, we
proceed in two stages. First we use established theory and a mixture of
observation, intuition and phantasy to set up a theory of how things work.
Then we (or more likely our friends) make a determined effort to falsify our
theory by finding a measurement or observation that is inconsistent with the
theory. In most cases this effort succeeds fairly quickly, and we have to
construct a new theory to continue the game. Occasionally resolute attempts
to falsify the theory fail, and people gain confidence in the accuracy of
its predictions. The theory is then considered established and becomes part
of the infrastructure that is called upon in the first phase of the
scientific process.

Sitting through this meeting, the conviction has grown on me that the Cold
Dark Matter (CDM) theory has now reached the point at which it should be
admitted as a Candidate Member, to the Academy of Established Theories, so
that it can sit alongside the established theories of Maxwell, Einstein and
Heisenberg and be used as a standard tool in the construction of new
theories.

I start by summarizing the status of the CDM and MOND theories, followed by
lists of headaches that the protagonists on each side have to confront, all
interlarded with my current views on relevant issues.  I finish with a `to
do' list. Since space is limited, I mostly rely on other contributors to the
meeting for citations to individual papers: the word {\sc Bloggs} is
shorthand for ``Bloggs, this volume and references therein.''

\section{Cold Dark Matter's Resum\'e}

It is now about twenty years since cosmology became dominated by
the theory that gravitational clustering is driven by CDM. This theory was
honed by interaction with observations of galaxy clustering, the Ly$\alpha$
forest, the CMB and type Ia supernovae. At this meeting several speakers
have concluded that astronomical data, especially those connected with
gravitational lensing, agree with the predictions of the CDM theory.
Recently the theory was rigorously tested by the WMAP satellite and passed
brilliantly ({\sc Spergel}). It will be further tested when data for
additional years have been analysed. It is a theory with a significant
hinterland in high-energy physics and as such can be tested in terrestrial
experiments: attempts to detect the particles of cosmic CDM as they pass
through the Earth are currently entering an exciting phase ({\sc Sellwood}),
and from around 2007 the LHC at CERN is expected to probe aspects of the
underlying physics, which probably involves supersymmetry.

In the last few years it has been widely argued that there are conflicts
between the predictions of CDM and observations of the internal structure of
galaxies of various types. Much of this meeting has been taken up with
discussion of these important questions. I argue below that these
problems probably reflect difficulties in correctly deducing the predictions
of the theory, rather than problems with the underlying physics.

\section{Modified Newtonian Gravity's Resum\'e}

Milgrom made an important contribution to our subject by having the courage
suggest that flat rotation curves reflected a failure of general relativity
rather than the existence of vast quantities of unseen matter (Milgrom
1983). In a series of lucid and painstaking papers he explored in
considerable depth this possibility, which he dubbed MOND. The essential
ingredient of MOND is the assertion that there is a characteristic
acceleration, $a_0\sim cH_0/2\pi$, below which
Newtonian gravity fails: when Newton predicts a gravitational acceleration
$g_{\rm N}\la a_0$, the actual acceleration is $g_{\rm M}=\sqrt{g_{\rm
N}a_0}$. 

Physicists are generally reluctant to give MOND the time of day because,
despite the best efforts of Bekenstein \& Milgrom (1984), a Lorentz
covariant theory that has MOND for a Newtonian limit has not come to light.
Astronomers tend to be more open-minded, and several studies have
demonstrated that Milgrom's MOND fitting formula enables one to predict the
rotation curves of galaxies from their light profiles ({\sc Sancisi; Sellwood};
Sanders \& McGaugh 2002).  What makes these fits impressive is that the
mass-to-light ratios $\Upsilon_C$ that they assign to each galaxy vary with
waveband $C$ and with galaxy type very much as one would expect on
astrophysical grounds if the galaxies contained only stars and the observed
gas ({\sc Bosma; de Jong}). These studies have
clearly established that a characteristic acceleration $a_0$ is involved in
the relation between galaxy photometry and dynamics. This fact is at least 
as important for our understanding of galaxy structure and formation as the
Tully-Fisher and Faber-Jackson relations. 

There now seems to be little doubt that Einstein's equations give an
adequate account of the data only when a significant cosmological constant
$\Lambda$ appears in them. Physically, the non-zero value of $\Lambda$
implies that the vacuum is a medium in which the energy density is non-zero,
and $\Lambda$ quantifies the density of this `dark energy'.\footnote{This is a
poor name because dark energy comes in two forms. The simplest form is `dark
matter' (DM). This exerts negligible pressure. The form quantified by $\Lambda$
is associated with a large negative pressure, i.e., a tension. So the
physical quantity associated with $\Lambda$ ought to be called `dark
tension' rather than dark energy.} The existence of dark energy strongly
suggests that our understanding of the way the vacuum responds to weak
stimuli is seriously in error. Milgrom and others have suggested that there
is a connection between the phenomenon of dark energy and the breakdown of
the predictions of standard gravity when small accelerations are involved in
the sense that both phenomena reflect differences between what actually
happens and the predictions of Einstein's theory with $\Lambda\sim3(a_0/c)^2$ set to
zero (Milgrom 2002). 

I find extremely seductive the idea that a better understanding of the
nature of the vacuum would simultaneously explain the requirement for
non-zero $\Lambda$ and modify the way that particles moved in weak
gravitational fields. MOND is a proposal for what the low-energy predictions
of the correct dynamical model of the vacuum would look like. So while MOND
itself could never become a member of the Academy of Established Theories,
its parent theory most certainly could.

Thus CDM and MOND do not compete on equal terms: the former is a natural
outgrowth of established physics, whilst the latter is a thing apart, a
single visible peak of a mountain range that is otherwise enshrouded in
cloud; a peak that is connected to the known world in an unkown way. The
result is that in our efforts to understand how the Universe got to its
present state, CDM can be used to form hypotheses and make predictions in a
way that MOND cannot.

\section{Headaches for MOND}

\subsection{Falling circular-speed curves} 

We heard that the Planetary Nebula Spectrograph on the William Herschel
telescope has measured velocity dispersions in the outer parts $R\ga2R_{\rm
e}$ of three intermediate-luminosity elliptical galaxies and find them all
to fall in Keplerian fashion ({\sc Romanowsky}). These data do not
absolutely require the circular speeds of these galaxies to be Keplerian,
but this is the most plausible interpretation of the data. Only two other
intermediate-luminosity elliptical galaxies have had their stellar
kinematics probed at $R\ga 2R_{\rm e}$ and only one of these (NGC 2434; Rix
et al., 1997) shows evidence for DM.  Milgrom \& Sanders (2003)
point out that these very high surface-brightness galaxies generate
accelerations around $R_{\rm e}$ that are an unusually large multiple of
$a_0$.  Consequently, MOND is expected to push the the circular-speed curve
above Newton's prediction only after a significant section of Keplerian
fall. Hence, the work reported by {\sc Romanowsky} is perhaps a triumph for
MOND.

\subsection{Non-universality of $a_0$}

The value of $a_0$ required to banish DM from rich clusters of galaxies is
about a factor of two larger than that fitted to data for disk galaxies
(Sanders \& McGaugh 2002 and references therein). {\sc Gerhard} argued that
the highest-quality data for elliptical galaxies at $R\la2R_{\rm e}$ implies
that DM manifests itself at accelerations $a\sim10^{-9}\,{\rm m}\,{\rm
s}^{-2}$ that are about an order of magnitude larger than the value of $a_0$
that is inferred from the dynamics of disk galaxies (Gerhard et al.\ 2001).
However, this conclusion would appear to conflict with the subsequent
results from planetary nebulae, which demonstrate that in two of the
galaxies in the Gerhard et al.\ sample (NGC 3379 and NGC 4494), even 4 to 6
times further out than Gerhard et al.\ were able to go, there is no
convincing evidence of DM. Baes \& Dejonghe (2002) point out that scattering
by dust at large radii of photons emitted near the centre can generate the
kinematic signature of a DM halo when none is really present.
Perhaps this effect is significant at the largest radii reached by Gerhard
et al.

\subsection{Halo flattening}

Weak lensing studies, like studies of polar rings, suggest that the
gravitational potentials of disk galaxies are significantly flattened. Is
this a problem for MOND, as {\sc Hoekstra} argued? It was not clear to me
that those who answer `yes' have borne in mind the strange behaviour of
multipoles in MOND. Milgrom (1986) shows (i) that you cannot determine the
quadrupole of the potential from the only quadrupole of the mass
distribution, and (ii) that outside the body the quadrupole of the potential
decays as $r^{-\surd3}$, rather than $r^{-3}$ as in Newton's theory.
Moreover, weak lensing is generated by accelerations that are only a few percent
of $a_0$. To impart the velocity $\sim600\,{\rm km\,s}^{-1}$ of the Local Group with
respect to the CMB in a Hubble time, takes an acceleration $a\sim
cH_0/500=0.012a_0$. Hence a single galaxy is unlikely to dominate the
deflections probed by weak lensing, and non-circular isopotentials are not
unexpected in MOND.

\subsection{Merging galaxies}

A famous photograph by Schweizer (1982) left little doubt that the merger of
two disk galaxies of comparable mass yields an elliptical galaxy. The
photograph shows the two long tidal tails of NGC 7252, together with the
galaxy's nearly relaxed core. The timescale on which the tidal tails evolve
will be the same regardless of whether galactic rotation curves are kept
flat by massive dark halos or modification of the law of gravity, because
speed and distance are both fixed by the observations. Thus the time since
the galaxies came into close contact is known.  Schweizer showed the
brightness distribution of the core obeys the $R^{1/4}$ law that is
characteristic of elliptical galaxies. Thus the nuclei of the two galaxies
have already completely merged. Simulations show that the nuclei can only
spiral together in the time available if they can effectively surrender
their energy and angular momentum to dark halos ({\sc Carignan}; Barnes 1988). If we banish the halos by modifying the law of gravity,
the galactic nuclei take much longer to merge because the vacuum cannot
relieve them of their energy and angular momentum.

\section {CDM and Galaxy Formation}

History shows that even when we have the correct theory, prediction can be
very difficult, even problematic: Newton published the {\it Principia} in
1686 but it was not until 1882 that Newcomb gave the now accepted value for
the Newtonian precession of the perihelion of Mercury, and it is only ten
years ago that Wisdom and others showed that there is significant chaos in
the solar system; Euler wrote down his equation for hydrodynamics in 1755,
but even in the sixth edition of his magisterial survey {\it Hydrodynamics}
Lamb (1932) displayed a very incomplete understanding of shock fronts and in
Art 284 ridiculed the Hugoniot jump condition.  It is only in the last
quarter century, with the development of chaos theory, that we have begun to
understand the devastating impact that the non-linearity of Euler's
equations has on the equations' predictive power.

CDM is a theory of the invisible. Apart from experiments designed to detect
wimps in the laboratory and gravitational lensing work, tests of the CDM
theory centre on the effect that CDM has on baryons. Hence galaxy-scale {\it
tests\/} of CDM are inextricable entwined with the theory of galaxy
formation, which nobody imagines is well understood. Because baryons
interact with each other and electromagnetic radiation in tremendously
complex ways, much cosmology is done with just CDM and dark energy.  We do
however now think that the mean density of baryons is $\sim{1\over5}$ that
of DM, so the exclusion of baryons from the calculations is far
from safe even on large scales. And baryons contribute much more heavily to
the density precisely at the locations, in or near galaxies, where we make
observations. So CDM-only simulations are, prima facie, unlikely to be
reliable guides to the configuration of DM in and around galaxies.
It's not going to be easy to wring reliable predictions for galaxy-scale
phenomena from the CDM model. Just as the fact that it is difficult to
calculate the long-term dynamics of the solar system is no criticism of
Newtonian dynamics, so the difficulty of predicting the DM
distribution expected in the Milky Way is no criticism of the CDM theory.

Pure CDM simulations and galaxy-formation models tend to be done by the same
workers in the same computers, but they are sharply different activities.
The dynamics of pure CDM is reasonably clear-cut and well understood. We can
feel confidence in the predictions it yields for large-scale phenomena, to
which baryons contribute only modestly. These predictions are in excellent
agreement with observation ({\sc Spergel}). I have the impression that the
predictions of pure CDM theory are in conflict with observation only where
baryons either are, or likely have been, dynamically important. During this
meeting I have come to the conclusion that we should treat these conflicts
not as falsifications of CDM theory, but as pointers to failings in our
understanding of galaxy formation. We are not even
sure how baryons should be accumulated at the centres of DM halos
in the purely dissipative case: the widely adopted prescription of
conserving the adiabatic invariants of the DM as the potential gets
deeper is probably quite wrong in the likely case that baryons accumulate
through a series of mergers ({\sc Primack}).

There is abundant evidence that non-gravitational energy is important for the
dynamics of baryons that fall into DM halos of a wide range in mass, from
dwarf galaxies right up to groups and even clusters of galaxies. The
dynamics associated with this non-gravitational energy is currently no more
than speculation.

\section{Headaches for CDM}

Many of the difficulties that CDM faces on small scales derive from the
result that the centres of pure CDM halos are very cuspy.  Even in the case
of pure CDM, I believe we don't understand properly how cores form, or what
the smallest radius is at which we can trust the simulations. There are two
problems: (i) the simulations still don't reach as far down the mass
function as the observations do, and (ii) we have an inadequate
understanding of the role of discreteness effects. In the simulations the
latter are important at early times when the Zel'dovich waves break, and
they are distinct from the effects of two-body relaxation.\footnote{I used
to worry about the latter, but I feel that Binney \& Knebe (2002) showed
that two-body relaxation is unimportant -- see however Diemand et al.~(2003)
for a different point of view.} However, through the cries and smoke of
battle, something like a consensus seems to be emerging from the groups that
do large-scale simulations of gravitational clustering ({\sc Navarro}): in a
pure-CDM halo the density profile rises in to the smallest resolvable radii
with a slope $\alpha=-{\rm d}\log\rho/{\rm d}\ln r$ that gradually lessens
inwards, but is still $\ga1$ when last believable. The density profile of an
individual halo depends weakly or not at all on the power spectrum from
which the simulation starts (Knebe et al.\ 2002). 

\subsection{LSB galaxies}

Stimulated by this prediction, observers have invested significant effort in
determining the density profiles at the centres of the most DM
dominated galaxies, since these clearly provide the cleanest tests. Although
the controversy has yet to die, my impression is that they have shown that
Low Surface Brightness (LSB) galaxies are unlikely to have such cuspy
central mass profiles as the pure CDM simulations predict ({\sc Bolatto,
Bosma, de Blok, Gentile, Mateo, Swaters, Trott}). Dwarf spheroidal galaxies
show the highest degree of DM domination ({\sc Wilkinson}), and the
conclusion of Kleyna et al.~(2003) that the UMi dwarf has a nearly harmonic
core is certainly a headache for CDM, but Draco does have a cuspy potential
(Kleyna et al.~2002). It seems that in many, perhaps all LSB galaxies, the
potentials are less cuspy than pure-CDM theory predicts. Will this problem
be resolved at some future date by correctly adding baryons to the CDM
simulations?  The resolution will be hardest for the dSph galaxies, because
their present baryon content is negligible. However, these are precisely the
galaxies that are expected to lose most gas during galaxy formation, so it
is not clear that their baryon contents were always negligible.

\subsection{Halo substructure}

In the last few years the worry has been abroad that DM halos are predicted
to contain more substructure than the observed number of satellite galaxies
suggests exists. It seems that this problem may have gone away: the number
of objects predicted is a steep function of the sub-halo's peak circular
speed. Mapping this circular speed onto quantities observed for satellites,
such as luminosity or central velocity dispersion, is non-trivial. It is
now argued that when this mapping is done correctly, the CDM model correctly
predicts the observed number of satellite galaxies ({\sc Primack}).
Moreover, in one interpretation of phenomena associated with strong
gravitational lensing, a high level of halo substructure is inferred ({\sc
Schneider, Mao} and below).

\subsection{HSB galaxies}

Many arguments now indicate that the centres of high surface brightness
(HSB) galaxies are baryon-dominated:

\begin{itemize}

\item Gerhard et al.\ (2001) find  that  the dynamics of elliptical galaxies is
well explained by assuming that DM makes a negligible contribution
to their central mass densities. The mass-to-light ratios then required are
in the range expected on astrophysical grounds and vary with the colour of
the galaxy in the expected way ({\sc Gerhard}). Significant contributions to
the mass density from DM would spoil this picture.

\item The shapes of rotation curves are intimately connected to the
underlying stellar light profile ({\sc Sancisi}). The rotation curves of
early-type disk galaxies peak at extremely small radii and must be dominated
by stellar mass over most of the radial range probed ({\sc Noordermeer}).

\item In disk galaxies we more often than not
see a fast bar (one that extends to $\sim0.8$ of its corotation radius).
The timescale for this to surrender most of its angular momentum to a
slowly-rotating, embedding dark halo is $f\sim{1\over2}(x+1/x)$ times the dynamical
time, where $x= \rho_{\rm bar}/\rho_{\rm DM}$ ({\sc Athanassoula}). Since
the high frequency of bars at the centres of galaxies implies that $f$ is
quite large, $x$ must be far from unity.

\item The  non-axisymmetric velocities that a bar drives in surrounding gas
are proportional to the mass of the bar, while the mass of the halo is
limited by the overall rotation curve and the masses of bar and disk.  Near
maximal disks are required to generate velocities as high as those observed
in some external galaxies ({\sc Bosma, Weiner}). Similarly,  the Milky Way's
`forbidden' velocity features, such as the $3\,$kpc arm in the $(l,v)$ plots
for CO and HI, require all available mass in the inner few
kiloparsecs to be associated with the bar and the disk (Bissantz, Englmaier
\& Gerhard, 2003; Fux, 1999).

\item We can weigh the Milky Way's disk near the Sun and all recent
investigators conclude that its entire surface density of $\sim41\,{\rm
M}_\odot\,{\rm pc}^{-2}$ can be accounted by known stars and interstellar
gas ({\sc van Altena}; Cr\'ez\'e et al.\ 1998; Holmberg \& Flynn 2000;
Olling \& Merrifield 2001). Given that the total column density within
$1.1\,$kpc of the plane is only $71\,{\rm M}_\odot\,{\rm pc}^{-2}$ (Kuijken
\& Gilmore, 1991), one can show that even as far out as the solar
neighbourhood ($R_0/R_{\rm d}\ga2.7$), DM makes a smaller
contribution than stars to the local circular speed (Binney \& Evans, 2001).

\item The microlensing optical depth to the Galactic centre has to be due to
stars. Even the much reduced current values ($\tau_6\equiv10^6\tau\sim1.5$:
Popowski et al.\ 2001; Alfonso et al.\ 2003) can be explained only if all
the mass that the circular-speed curve can accommodate in the central few
kiloparsecs is invested in stars (Bissantz \& Gerhard 2002).

\end{itemize}

I'd like to abuse my privilege of having the last word to contribute to the
controversy about the implications of measurements of $\tau_6$. I coauthored
two papers (Binney, Bissantz \& Gerhard 2000; Binney \& Evans 2001) that
used an upper limit on the optical depth that can be achieved with a given
mass when it is distributed smoothly around an ellipse. In the first paper
we argued that the values, $\tau_6\sim3$, that were then current were
physically impossible.  Obligingly, Popowski et al.\ (2001) shortly
afterwards reduced to $2\pm0.4$ the most reliable value of $\tau_6$, that
for red clump stars, which are bright enough for blending not to be a
problem. In the second paper we showed that even the optical depth of
Popowski et al.\ places a tight constraint on the power index $\alpha$ of
the dark halo's density profile between the Sun and the centre because we
know the local density of DM, and $\tau_6$ limits the density of
DM at the centre. 

Recently the EROS collaboration have reported $\tau_6=1.08\pm0.3$ (Alfonso
et al.\ 2003) and at this meeting Merrifield has argued that flattening the
dark halo to axis ratio $q=0.8$ allows $\alpha$ to be as large as unity
without lowering $\tau_6$ below its likely value. Hence I hear it said on
all sides that the Galactic microlensing data are now compatible with an NFW
halo.

This conclusion is very wrong! What has been lost in the debate is how {\it
absurd\/} the distributions of stars considered in our two papers are.
These distributions are the ones that maximize the optical depth along
$b\sim-3.8^\circ$ for a given mass subject only to the constraints (i) that
the density decreases exponentially with distance $z$ from the plane, with a
scale-height $z_0(R)$, (ii) that the matter distribution is elliptical, with
unconstrained ellipticity and a principal axis inclined at $20^\circ$ to the
Sun-centre line, and (iii) that the disk's radial density profile is
exponential. The resulting scale heights $z_0$ increase linearly with
distance $D$ from the Sun so that $z_0/D=|b|$, the latitude at which the
value of $\tau_6$ is set. The Galaxy's stars are not distributed in this
way! We used these absurd models to make the point that with the old data
there was no way in which the lenses could be distributed through the Galaxy
to achieve the required optical depth. Now that the microlensing
measurements have been refined and are producing plausible values, it is
time to add the constraint that the lenses, which are almost certainly
ordinary stars, are distributed like the starlight. Since this distribution
does not maximize the optical depth along a particular line of sight from
the Sun, it produces a smaller optical depth per unit stellar mass than
Binney \& Evans assumed. The analysis of the COBE near-IR light distribution
by Bissantz \& Gerhard (2002) provides our best estimate of the distribution
of stars near the centre. If these stars contain {\it all\/} the mass that
is allowed in the centre, Bissantz \& Gerhard conclude that $\tau_6=1.27$. So
at the present time there is very little room for DM in the inner
few kiloparsecs, even with the new lower optical depths.

\section{Things to do}

I have argued that ab initio prediction of the DM distribution in
galaxies is hard because it involves solution of the full galaxy-formation
problem. If we accept this conclusion, we should concentrate on {\it
mapping\/} DM in galaxies, in an open frame of mind. This enterprise has two
arms: (i) use of all available dynamical probes (stars, X-ray gas, atomic
and molecular gas, gravitational micro- and macro-lensing) to map total
mass, while (ii) using the most sophisticated models of stellar populations
and interstellar gas to infer baryonic mass. As we have heard from many
speakers, arm (i) is rather fully developed. By contrast, arm (ii) requires
much more work, especially as regards the ISM. I feel the work described by
{\sc de Jong} is a valuable step along the right path.

We must go after cold molecular gas like bloodhounds, while avoiding a false
dichotomy between `missing mass is all in DM', and `missing mass is
all in the ISM'. In the LSB regions of galaxies it is natural that gas
should condense into very cold clouds that are hard to see in emission ({\sc
Pfenniger}).  Observations in the $21\,$cm line have long told us that
galaxies have HI disks that extend into extremely LSB regions. If the ISM in
these regions were several times more massive than its HI component, we could
understand a number of things:

\begin{itemize}

\item The existence of spiral structure in the extended disk around NGC 2915
({\sc Masset}).

\item Why polar-rings lie above the Tully-Fisher relation for normal
spirals, while their embedded disk galaxies lie on it ({\sc Arnaboldi}).

\item The existence of the baryonic Tully-Fisher relation ({\sc Combes}; Matthews et al.\
1998; McGaugh et al.\ 2000).

 \item The presence $\sim20\,$kpc from the centre of M31 of a surprising
amount of dust and star formation. If stars are forming, there must be
unobserved H$_2$ present, and  if the ISM has the low
metallicity expected so far out, the observed level of reddening is hard to
understand without a significant column of H$_2$ ({\sc Allen, Pfenniger}).

\end{itemize}

Data from the new X-ray observatories have yet to have a big impact on
studies of individual galaxies because they require extremely long exposure
times, and we need to be more sophisticated in our modelling of hot gas. In
particular, models need to include the spin of the body of trapped gas. I think
there is considerable scope for linking our knowledge of stellar populations
to the composition and global dynamics of the ISM.

We need to clarify the situation regarding DM in elliptical galaxies.
Studies of gravitational lensing suggest that at $\sim2R_{\rm e}$ both stars
and DM make significant contributions ({\sc Schneider, Schechter)}. The flux
ratios in the images of strongly lensed background objects require the mass
distribution in the lensing galaxy to be lumpy ({\sc Mao}). {\sc Schneider}
took the lumpiness to be sub-structure in the DM halo, such as that
predicted by the DM clustering simulations.  {\sc Schechter} argued that the
lumpiness of the stellar contribution to the mass distribution was
responsible to the observed flux ratios.  Observations of the variability of
narrow and broad emission lines in the background source will resolve this
issue within a few years and greatly clarify the distribution of DM in
elliptical galaxies. 

If we exclude cluster-centre galaxies, the dynamics of
stars provides at best weak evidence for DM: the data favour some DM at
$R\sim2R_{\rm e}$, but what there is does not prevent the circular speed
dropping in near Keplerian fashion at $R\ga2 R_{\rm e}$ ({\sc Gerhard,
Romanowsky}). Compact DM halos that yield falling rotation curves
have long been favoured by studies of tidal-tail formation and evolution
(Dubinski et al.\ 1999 \& references therein).
There must be a worry that the rather different picture
provided by the lensing data is biased: even if only a minority of
ellipticals have massive dark halos, nearly all the observed lenses will
belong to that minority. 

Last but by no means least we have to press on with developing our
understanding of how galaxies form. This is going to be a long job, but an
immensely worthwhile one. We'll probably crack it soonest if we accept CDM
as background theory.

\end{document}